\documentclass[pre,twocolumn,floatfix,superscriptaddress,showpacs]{revtex4-1}
\usepackage{amsmath,amssymb,latexsym} 
\usepackage{rotating,lipsum}

\usepackage{graphicx}
\usepackage{color}

\begin{document}

\newcommand{\udem}{D\'{e}partement de Physique, Universit\'{e}
de Montr\'{e}al, Montr\'{e}al, Qu\'{e}bec, Canada H3C~3J7}

\newcommand{\nrc}{National Research Council of Canada, Ottawa, Ontario.
 Canada K1A 0R6}

\title
{A study of the electronic and ionic structure, for competing
states of fully and partially ionized hydrogen, using
the neutral pseudo-atom method as well as a classical map
for the electron subsystem.
}

\author
{
 M.W.C. Dharma-wardana}
\email[Email address:\ ]{chandre.dharma@yahoo.ca}
\affiliation{\udem}
\affiliation{\nrc}

\begin{abstract}
Prof. Bonitz and his collaborators have made seminal contributions
 to the study of the uniform electron fluid and 
the electron-proton fluid, viz., hydrogen, in using {\it ab initio} simulations.
 as reflected in this festschrift.
Here we 
review the theoretical methods available for these systems where
traditional small-parameter methods fail. We use 
the neutral-pseudo atom (NPA) method, and a classical map for
quantum electrons to study hydrogen plasmas.
We show that {\it both} fully-ionized  and partially-ionized
hydrogen phases can exist with the same nominal density and temperature,
at pressures and temperatures of interest to  planetary physics.
The mean ionization $\bar{Z}$, pair-distribution functions, free energies,
pressures and conductivities are calculated for the competing phases.
 Here  $\bar{Z}$ is also a measure
of the miscibility of fully ionized and un-ionized hydrogen.
Recent studies using path-integral
 Monte Carlo methods, and $N$-atom Density Functional Theory (DFT) simulations
 have provided essential  structure data including the electron-electron
structure factor $S_{ee}(k)$ that enters into interpretation of X-ray
 Thomson scattering and other diagnostics. We show that  these structure data
  can be inexpensively evaluated using classical-map schemes for fully
ionized plasmas, and more generally, using  one-atom (average-atom)
  DFT methods for partially ionized systems.
\end{abstract}

%
\maketitle

{\bf Conflict of Interest} - The author declares no conflict of interest related to this research.\\

{\bf  Funding} - This research did not receive any specific grant or funding from any sources in the public, commercial, or not-for-profit sectors.\\

{\bf Data availability} - All data used in the paper are availabe within the paper graphically or otherwise, and also by reasonable request from the author.\\

{\bf ORICD}  0000-0001-8987-9071 \\

\section{Introduction} 
The advent of high-energy lasers and femto- and even atto-second
technologies~\cite{Corkum07}  have made it possible to create and study
matter under extreme conditions~\cite{Kremp99, Ng95}.
Prof. Michael Bonitz and his collaborators have made seminal
contributions to the {\it ab initio} study
 of such systems~\cite{BonitzBk16, BonitzPOP24, FilinovBonitz23},
 as reflected in this Festschrift.
 
Material systems under extreme densities $\bar{\rho}$,
and temperatures $T$,  with extreme pressures $P$
occur naturally in planetary interiors and astrophysical
 objects. They also occur as transient states that have to be probed
 on sub-nanosecond  timescales~\cite{Ng95,Poole24,Drake2018,Betti2016,
GaffneyHDP18,McBride-Si-19}. These involve
cutting-edge experiments on high-energy-density materials relevant
to experimental astrophysics~\cite{Drake2018}, 
fusion physics~\cite{Betti2016,GaffneyHDP18}
and even for nuclear-stockpile stewardship requirements. The Fermi energies
 $E_F$ of compressed matter are large, and although $T$ (where we use
 energy units)  may be nominally high, $\theta=T/E_F$ is of the order of
unity. Hence
 the name ``warm-dense matter'' (WDM) has
been used for such highly-correlated energy-dense matter.

The number of free-electrons per ion, viz., $\bar{Z}$, the interaction potentials,
  static structure factors $S_{ab}(k)$ where $a, b$ = $e,i$
 (i.e., electrons, ions)  as
well as their dynamic analogues, e.g., $S_{ab}(k,\omega)$~\cite{HARB-DSF18} are
 relevant to the evaluation of the equation of
 state~\cite{eos95, GaffneyHDP18}, and
transport properties~\cite{eos95, Stanek24} of WDM. 
They appear in the analysis of 
experimental results from  X-ray Thomson scattering
 (XRTS)~\cite{Gregori03,GlenRed09,Doppner23,Dornheim25,CDWBe25}
and X-ray diffraction~\cite{Poole24,cdwSi20,McBride-Si-19,CDW-Pool25}.
The theoretical methods available for the study
of WDM hydrogen and the uniform electron fluid (UEF)
 include the following.
\begin{enumerate}
\item  $N$-atom density-functional calculations,
coupled with molecular dynamics (DFT-MD) where $N$ nuclei
and the corresponding number of electrons are treated using Kohn-Sham-Mermin
density functional theory (DFT). Here the
ion configurations are evolved using classical molecular dynamics (MD). 
This method is referred to as DFT-MD, or more frequently as QMD (quantum MD).
\item Path-Integral-Monte-Carlo (PIMC)
 methods. Here the partition function is evaluated using Feynman trajectories,
and works most efficiently at high-$T$ where the trajectories are close to
classical trajectories.
\item  Average-atom (AA) methods. In the specific Average-Atom model that we use,
 the
$N$-nuclei problem is reduced to a one-body (i.e., one-ion) problem using an ion-ion XC functional
to construct a neutral-pseudo-atom (NPA) model; this is an exact DFT
procedure, to the extent that the e-e and ion-ion XC functionals are accurate.
\item Methods based on using classical potentials that include quantum effects.
Here we use the Classical-Map (CMap) procedure of Dharma-wardana and Perrot
to accurately map the quantum-electron system at temperature $T$ 
to a classical Coulomb fluid at an appropriate temperature $T_{cf}$.
\end{enumerate}.

We present comparisons of $S_{ab}(k)$,  and other relevant quantities from
these methods, for the case of the UEF, and for hydrogen plasmas. The
case of two competing phases of hydrogen, viz., fully and partially ionized
hydrogen is studied in detail.

\section{Review of theoretical methods}
We review the features of the theoretical methods enumerated above
in greater detail below, partly with an eye to understanding
why these methods may not be successful in exposing phase transitions.
Here we note that various QMD calculations over several decades failed
to reveal a phase transition in $l$-carbon~\cite{Hull20} near the diamond density;
it required NPA calculations and QMD-SCAN~\cite{cdw-carb22}
 calculations to bring it to light. 

We use atomic units with $\hbar = m_e = |e| = 1$, while
 $T$ will usually be given in energy units of eV (1 eV = 11,604K). We also define
the electron- and ion-  Wigner-Seitz radii $r_s = [3/(4\pi\bar{n})]^{1/3}, 
r_{ws} = [3/(4\pi\bar{\rho})]^{1/3}$, where $\bar{n}$ is the average
free-electron density in the plasma, while $\bar{\rho}$ is the average
ion density. The number of free electrons per ion
(experimentally measured using a Langmuir probe in low-$T$ plasmas) is
$\bar{Z}$, with  $\bar{Z} = \bar{n}/\bar{\rho}$. The Fermi wavevector
$k_F = (2\pi^2\bar{n})^{1/3}$ and the Fermi energy $E_F = k_F^2/2$
are important scales of energy and momentum for the electron subsystem.
Note that the review of Bonitz et al.~\cite{BonitzPOP24} seems to use $r_s$ 
for  $r_{\rm ws}$, even for partially-ionized hydrogen, without distinction.

\subsection{$N$-atom DFT simulations with MD evolution of ions}
\label{DFT-MD.sec}
$N$-atom Density Functional Theory (DFT) uses $N$ nuclei, located
at $\vec{R}_I, I = 1,..,N$ and the corresponding number of electrons.
usually with $N\sim 64-256$. The multi-centered 
electron density $n(\vec{r};\vec{R}_I)$
is calculated using Kohn-Sham theory in the fixed potential of
the ions treated as in a periodic (crystalline) potential.  
In effect, the Hohenberg-Kohn stationary condition on the
 one-body electron density leads to the Kohn-Sham equation for
a set of non-interacting electrons at the interacting density.
\begin{equation}
\label{eeEulerLag.eqn}
\frac{\delta F([n(r)];\vec{R}_I)}{\delta n(r)}=0
\end{equation}
Thus  the interacting plasma is mapped into a Lorentz plasma
in QMD with the ions held fixed. Only the electron many-body problem
 is simplified in this manner, with the ion-ion many-body 
problem left intact. 
So, the many-ion configuration $\vec{R}_I$ is evolved to equilibrium 
 using classical molecular-dynamics (MD)
to obtain a statistical thermal average for the multi-centered
one-body density. This procedure is implemented in many
commercially available computational packages~\cite{VASP, ABINIT}, where 
a choice of pseudopotentials, and $T$ = 0 XC-functionals are available.
Currently  commercially available packages do not provide finite-$T$
functionals. However, implementations of
finite-$T$ codes and calculations (e.g., ~\cite{KarasievHu19,Ramakrishna20}) have been reported
for QMD, while Ref.~\cite{KarasievQE14} presents orbital-free DFT applications.

 The many-ion problem
 in QMD has to be treated ``head-on'' via an $N$-atom classical simulation. The MD
 procedure yields the ion-ion pair distribution function (PDF),
viz., $g_{ii}(r)$, and the corresponding $S_{ii}(k)$,
 but further computationally costly manipulations are needed to obtain
 the Kubo-Greenwood response functions required
for the determination of the electrical and thermal
conductivities $\sigma$ and $\kappa$~\cite{Recou05}. Similarly, further
manipulations are needed~\cite{MoldHyd25} to determine $S_{ee}(k)$ that
enters into  XRT spectra~\cite{Plage-XRTS15,xrt-Harb16,CDW-Pool25}.

The mean number of free electrons per ion, viz.,
the mean ionization $\bar{Z}$ is not easily obtained from
such $N$-atom QMD calculations, leading to several
differing estimates of $\bar{Z}$~\cite{SternZbar07,BethkenZbar20}.
This method, variously referred to as DFT-MD, KS-MD (Kohn-Sham MD)
and QMD (quantum molecular-dynamics)~\cite{GaffneyHDP18} will be
referred to here as QMD due to its brevity and wider usage.
QMD is an established ``workhorse'' although it has several major
 limitations: 

1. its accuracy decisively depends
 on the electronic exchange-correlation (XC) functional. 
Thus, estimates of the conductivities ($\sigma,\kappa$) using QMD
may differ by a factor of two depending on the selected
 XC-functional~\cite{Stanek24}.

2. Given an  exact XC-functional, the Hohenberg-Kohn-Mermin
 theorem of DFT  promises to give the
total free energy $F$ (or the ground-state energy $E$ at $T$ = 0)
 and the one-body distributions $n(r),\rho(r)$ of the electrons
 and ions. But the theory does not provide valid excitation
 energies or electron eigenfunctions. The
Kohn-Sham method maps the interacting many-electron system
 to a non-interacting system and provides the one-electron
 energies and eigenfunctions of that {\it fictitious} electron
 system. However, they are still a good approximation in many cases;
  but the bands gaps etc., of the non-interacting electron system
obtained by the Kohn-Sham method need further corrections if
 they are to represent the actual physical system. For instance, DFT
calculations of solid Ge and many other semiconductors
return a metallic solid with no band gap or greatly reduced
bandgaps~\cite{Sham85,HybLouie85}.

Most implementations of the Kohn-Sham method do not correct the
 Kohn-Sham states for self-interaction effects~\cite{PerdZung81,cdw-Carbon10E6-21},
 or for the discontinuities in the XC-functional across
 band gaps~\cite{Kohn86}. Thus, when the
Kohn-Sham eigenstates and eigenfunctions are directly used for,
say, the calculation of the mean ionization $\bar{Z}$,
results with varying degrees of inaccuracy
may be obtained~\cite{cdw-Carbon10E6-21,Fauss21,BethkenZbar20}.
These effects may be substantially corrected if the DFT-MD calculation is
augmented with a GW-quasiparticle  calculation~\cite{HybLouie85,PDW-Dyson84}
but the computational costs becomes prohibitive, especially for finite-$T$
applications.
 
3. $N$-atom DFT coupled with MD (i.e., QMD), with $N\sim 100-500$
 at finite-$T$ is computationally extremely expensive even without
 GW-quasiparticle corrections because the basis sets needed for the
 finite-$T$ calculations increase rapidly for finite $T$.
 For instance, the recent
 analysis  of XRTS spectra of highly  compressed
 Be~\cite{Doppner23,CDWBe25,Dornheim25}  using QMD would take months
 of computation, and hence many laboratories choose to remain outside
such warm-dense matter studies. 

Accurate orbital-free DFT methods have been
 a long-sought goal for speeding up these $N$-atom DFT calculations.
 However, while considerable progress has been made in developing
 kinetic-energy functionals~\cite{Trickey23}, their accuracies remain
 well below Kohn-Sham calculations; they also
 do not provide any one-electron functions that are convenient
 intermediaries in calculating many physical properties of WDM systems.

4. The structure factors $S_{ab}(k)$ are available only for $k > k_0$ where $k_0$
depends inversely on the linear dimension of the simulation box. For instance,
Moldabekov et al.~\cite{MoldHyd25} do not report  data for $k < 1.15$ \AA$^{-1}$
in their  DFT-MD studies of hydrogen at 0.33 g/cc. The issue of obtaining
the small-$k$ limit of structure data is discussed further in Sec.~\ref{kto0.sec}.

\subsection{Path-integral and Monte-Carlo methods}
\label{pimc.sec}
A quasi-exact method for quantum simulations is afforded by  
path-integral Monte Carlo (PIMC)
 simulations, if not for  errors arising from the
 Fermion-Sign Problem (FSP) that increase rapidly as $T/E_F$ decreases.
~\cite{Loh90,Troyer05,Dornheim2019,DornheimPIMC24}. Bonitz et al.~\cite{BonitzPOP24}
state that ``When considering hydrogen at temperatures well below the electronic -
 degeneracy temperature,..., fermionic PIMC
(FPIMC) becomes highly inefficient because of the FSP".
Approximate methods for dealing with the FSA have been developed,
and applications to WDM studies, e.g.,~\cite{driver12,ZhangMilitzeCH-18}
and for the calculation of finite-$T$ XC-functionals~\cite{Brown2013,GDS17}
 are well known. We refer to all these approximate  methods 
as PIMC for brevity.

We use in this study the PIMC results at $r_s =$ 2, and 3.32 for warm dense
hydrogen provided by Moldabekov et al.~\cite{MoldHyd25} to compare
with similar results from an approach based on a  classical-map of the
quantum electron system where a statistically equivalent classical Coulomb gas is
studied. 

\subsection{Limitations of PIMC and QMD}
Given that both PIMC and $N$-atom DFT-MD (i.e., QMD)
are computationally very demanding, and may take months of calculation for
analyzing current state-of-the-art experiments, accurate first-principles
 methods that are also computationally very efficient are badly needed.
These methods do not yield emergent properties of complex systems,
e.g., H-atoms, or excitons as the description is only
in terms of electrons and nuclei.
Capability for fast prediction would
 enable the experimentalist to optimize the diagnostics and settings
 ``on-the-fly", i.e., while the experiment is in  progress. For instance,
we may consider the prediction of the XRT spectra, or, at the more
 fundamental level, the prediction of  pair-potentials $V_{ii}(r)$, structure
factors $S_{ee}(k)$ within computational
times of a few seconds as an objective for theory development.

\subsection{Average-atom models}
\label{AA.sec}
A serious bottleneck in $N$-atom DFT is the need
to calculate Kohn-Sham states for tens of thousands of ionic configurations
in resolving the ion-ion many-body problem, when this can actually be
reduced to a {\it single} Kohn-Sham calculation. That is, the ion-ion many-body
 problem can also
be rigorously reduced to a one-body problem in DFT using an appropriate ion-ion
XC functional~\cite{DWP82}, just as the electron-electron problem is
reduced to an effective one-electron problem using an e-e XC functional. 
In effect, instead of using
just Eq.~\ref{eeEulerLag.eqn} dealing with only electrons, we consider
two coupled equations. The first of these is for the electron subsystem
with the one-body density $n(r)$, and the second deals with the
 ion one-body density $\rho(r)$. We are considering a uniform system
 with the mean electron density $\bar{n}$ and a mean ion density $\bar{\rho}$.
The resulting analysis  leads to the construction of a single representative
atom of a single nucleus placed in the appropriate plasma medium. It is
referred to here as the neutral-pseudo-atom (NPA) model~\cite{DWP82}. It is a
rigorous DFT model where  e-e and ion-ion
many body effects are reduced to  suitable one-body XC-functionals, while
certain electron-ion correlations~\cite{Furutani90} have been neglected.
 The latter was the
main criticism leveled by Chihara~\cite{ChiharaNPA91} against the NPA model
of Dharma-wardana and Perrot~\cite{DWP82}. Details of computationally convenient
implementations of the NPA have been given
 in many past publications~\cite{eos95,Pe-Be,Hungary16}.

While the NPA begins as an all electron calculation,
it constructs, internally, many intermediate quantities to simplify
 the calculation and enlighten the physical picture.
 These involve defining an average ionization $\bar{Z}$, 
ion-electron pseudopotentials, T-matrices, as well as ion-ion pair potentials
and pair distribution functions.
 All intermediate quantities
  other than XC functionals are directly computed from
 the results of the  Kohn-Sham  calculations. So,  given the material
 density $\bar{\rho}$,  the  nuclear charge $Z_n$,  and the temperature $T$,
 the only  quantities needed in the NPA from an external theory are
 the XC-potentials. 

In the NPA, $\bar{Z}$ is evaluated from the finite-$T$ Friedel sum rule,
 using the phase shifts of the continuum electrons (see the Appendix).
 These continuum
 states are not significantly subject to self-interaction errors or
 XC-discontinuities. The phase shifts are calculated from the behaviour
of the scattered wavefunctions in the asymptotic region, i.e., where
interactions are negligible, and when the fictitious 
Kohn-Sham states and the physical (Dyson) states~\cite{PDW-Dyson84}
 asymptotically agree.

 Also, the evaluation of the phase shifts is made at the minimum of
 the total free energy  $F([n],[\rho])$ for variations in the one-body
 densities $n,\rho$.  This means $\bar{Z}$ corresponds to
a full Saha-type calculation inclusive of all the many-body corrections
automatically  included via XC-functionals. This ``automatic'' free-energy
 minimization picture that arises in an NPA calculation
 is presented in some detail in Ref.~\cite{eos95}.

That is, the  NPA does not attempt to impose Yukawa- or Debye-H\"{u}kel
models, Lenard-Jones models, bond-order models etc., taken from outside.
 However, the ion-electron  pseudopotentials
 obtained from the NPA can be {\it post facto} further
re-parametrized for convenience and transferability, e.g.,
as modified Heine-Abarankov pseudopotentials~\cite{ELR98}.
The ion-ion pair-potentials can be represented by Yukawa
forms augmented by a Friedel-type oscillatory tail~\cite{DW-yuk22}.
The neglect of these oscillatory tails prevents the correct
prediction of subtle liquid-liquid phase transitions, or, for instance,
 Martensitic-type transitions found in many materials.

It has been claimed that the NPA, or average-atom
models cannot correctly simulate systems such as liquid carbon
at low temperatures due to the existence of complex
covalent bounds in them~\cite{HamelCH12,whitley15}.
 We have demonstrated in previous publications
that the ion-ion PDFs of complex
 fluids of  materials such as C, Si,
obtained using $N$-atom DFT-MD can also be obtained using the NPA,
in close agreement with the QMD results, even close to the
melting line. Since the NPA is
based on static DFT, the calculated PDFs, $S_{ab}(k)$ are
long-time averages where transient bonding effects
demonstrated in $N$-Atom DFT simulations, e.g., in Ref.~\cite{HamelCH12}
 are of no consequence to thermodynamic and static quantities
such as the PDFs of uniform fluids.

\subsection{Accessing the small-$k$ limit}
\label{kto0.sec}
An advantage of average-atoms models such as the NPA over PIMC and 
QMD is that the structure factors $S_{ab}(k)$ are available
for the whole range of $k$ values including the $k\to 0$.
 For instance, a QMD calculation using 
64 atoms corresponds to a linear dimension of only
4 atoms, i.e., only a single nearest-neighbour to any
 ``central'' atom is treated, and
the accessible $k$ is limited to values larger than $\pi/L$.
Furthermore, moving an atom in the simulation causes a
density fluctuation of the order of $1/N$ and may washout
any distinction between metastable states.
Similarly, Rayleigh-weight calculations of XRT spectra require results
at low scattering wavevectors where many-body effects become
more prominent; but such calculations for small-$k$
are extremely expensive using PIMC, QMC or QMD.
      
This problem of predicting small-$k$ limits of structure data
cannot be side-stepped by using ``machine-learnt'' pair potentials (MLPs),
unless the MLPs have been trained on expensive
PIMC, QMC and QMD data that contain information on the small-$k$ region.
 If the theoretical data were supplemented
with empirical data, then the results are not {\it ab initio}.
Furthermore, while there exists a theorem that validates the search for an
universal XC-potential as a functional of the density, there is no
``existence theorem" that asserts that MLPs of a general nature exist.
The short-comings of such MLPs (which
are usually multi-centered functions dependent on many parameters with little
or no physical meaning), have been discussed recently, together with comparisons
from NPA pair-potentials~\cite{DW-yuk22,Stanek21,cdw-pop21}. The
 simple NPA approach will
remain very competitive even if such MLP-schemes could be widely implemented.

Furthermore, the classical-map approach discussed in Sec. V  provides an inexpensive
way of extending PIMC structure data to the small-$k$ region. This is illustrated
by our Cmap calculations for recovering the $S_{ee}(k)$ reported by Molabekov et al
for $r_s$ = 2, $T$ = 12.5 eV.

\subsection{Short-comings of the NPA model}
A theoretical critique of the NPA model has been given by Chihara~\cite{ChiharaNPA91}.
 However, applications of the
NPA to actual WDM calculations show that Chihara's  main objection (that a third XC 
functional to cover e-i exchange-correlation effects is lacking in the NPA) has no practical impact.

In our experience, the main short-comings of the currently implemented NPA code are as follows.\\
1. It is limited to uniform systems, or to well-defined crystalline systems. For instance,
while the current NPA code could reproduce the high-density branch of the Ganesh-Widom
 pressure curve~\cite{GaneshSiLPPT-09}
for $l$-Si at its melting point, the low-density metastable
supercooled-Si branch could not be demonstrated~\cite{cdwSi20,Remsing20}.  \\
2. The current NPA code  assumes a sufficient presence
 of free electrons to justify the use of certain linearizations, e.g.,
in constructing pseudopotentials or pair potentials used within the code. \\
3. As it is a DFT code, the NPA eigenfunctions, eigenvalues, and level occupations
are for the fictitious non-interacting Kohn-Sham electrons. The code does not implement
self-interaction corrections or the discontinuity in the XC-functional when the number
of electrons changes by an integral number~\cite{Kohn86}.\\
4. The many-body hamiltonian is rigorously replaced by a system consisting of
a uniform electron gas containing embedded ``hydrogen atoms'' of effective charge $\bar{Z}$
carrying a core of bound electron. NPA being a DFT theory,
 more complex associations like (H$_2$)$^+$ etc., 
being many-body effects, are not explicitly sorted out although their effects are
 included via XC-functionals. This may be viewed as a short-coming in some
 sense, while being a computational advantage of the theory! 
5. The finite-$T$ XC-functional implemented in the NPA code is that of  Perrot
 and Dharma-wardana(PDW-XC)~\cite{PDWXC}, based on the classical-map hyper-netted-chain model. 
This is usually not a practical limitation as
 structure factors, free energies, EOS  etc.,
 calculated using PDW-XC or the finite-$T$ GDS-XC by Groth et al.~\cite{GDS17} give
 very close results for typical WDM-test cases. This is established in Sec.~\ref{clm.sec} for
the hydrogen WDMs studied here.\\
6. A spin-density functional treatment is not available in the current NPA code, making it
unsuitable for applications to transition-metal systems at low-$T$.   
 
\subsection{The classical-map approach to WDM calculations}
Attempts to use classical ``statistical''
potentials that mimic quantum effects has a long history
~\cite{Filinov2004}. The classical-map  (Cmap) approach developed by
 the present author and Perrot~\cite{prl1}
was  motivated entirely by DFT ideas.

We noted that a fully non-local
 ion-ion XC-functional that is needed to reduce the ion-ion many-body problem could
be internally generated within the calculation itself, using classical
statistical mechanics. This means, given an accurate mapping of quantum electrons into a statistically
equivalent classical Coulomb fluid, then the fully non-local XC-functionals needed
for reducing the electron many-body problem could also be generated.
 Furthermore, the whole
calculation reduces to classical statistical mechanics, 
avoiding the heavier machinery of quantum-statistical
mechanics. 

The current state of the classical map (CMap) approach is that
it is quite successful in dealing with the
normal  uniform electron fluid (UEF)
at zero to finite $T$, at any electron
density $\bar{n}$ and spin polarization $\zeta$. The method
can be justified using DFT. Briefly, what has been
 established~\cite{PDWXC,prl1,prl2,SandipDufty13,
Bredow15,hug02,cdw-N-rep19,LiuWuCHNC14,lfc1-dw19}
 is that 
the PDFs of a classical fluid (consisting of classical electrons
 interacting via a 
Coulomb-like interaction) become equal to that of the
quantum UEF at the given $\bar{n}, T, \zeta$, if the
 temperature $T_{cf}$ of the
classical fluid is chosen such that the  XC-energy of the
classical Coulomb fluid is equal to the
XC- energy of the quantum fluid at the
corresponding $\bar{n},T,$ and $\zeta$.
 
At the time when the classical-map approach was developed,
finite-$T$  $F_{xc}$  data for the UEF
were not available. Hence the classical-fluid  temperature $T_{cf}$
at $T$ = 0 was evaluated using the $T$ = 0 XC-energy. This  was
named $T_q$, and the classical-fluid temperature
$T_{cf}$ at finite-$T$ was estimated using the ansatz
\begin{equation}
\label{tcf.eqn}
T_{cf}=\surd\left[T^2_q +T^2\right]
\end{equation}
Here $T_q$ depends only on the $r_s$ of the electron fluid. Denoting the
spin states as `$u, d$', the classical
Coulomb fluid at $T_q$ recovers the PDFs $g_{uu}(r), g_{ud}(r)$, and
$g_{dd}(r)$ at $T$ = 0 and for arbitrary spin polarizations $\zeta$. Here
we are concerned with the case $\zeta=0$, and more details of
CMap calculations may be found in Refs.~\cite{prl1,hug02,Bredow15}.  

In effect, since there
is no Hartree contribution in uniform fluids, the total free
energy is simply the XC-energy. DFT asserts that when two systems
modeled to have the same Hamiltonian have the same free energy,
 then their one-body distributions,
viz., $n(r) = \bar{n}g_{ee}(r)$ become equivalent. That is, the $g_{ee}(r)$
of the classical fluid becomes equivalent to that of the quantum fluid.

While the classical map is successful for mapping the quantum UEF, it
is of limited success with partially ionized hydrogen
 plasmas~\cite{Bredow15, cdw-N-rep19,LiuWuCHNC14,hug02},
or for more complex systems like aluminum plasmas.
Even in the fully ionized case, while the effective classical temperatures for
 the e-e and i-i interactions are evidently $T_{cf}$ and $T$,
 the temperature $T_{ei}$ of the electron-ion interaction is open to modeling.
 Such a temperature
 can be constructed via  the compressibility sum rule~\cite{Bredow15},
 or by demanding that $g_{ei}(r)$ be consistent with that of
an NPA calculation. 
 However, as with PIMC calculations, for sufficiently
high $T$, the CMap method can be used. That is, at high $T$, the differences
among $T$, $T_{cf}$, or $T_{ei}$  become small and the
 method begins to work well. 

In this study we present CMap calculations of
$S_{ab}(k)$ for fully ionized hydrogen for comparison
with PIMC calculations where available. 
Such comparison are a necessary step for improving the
classical-map approach and other computationally
fast methods.

\section{The degree of ionization $\bar{Z}$ for Warm dense Hydrogen}
\label{ZbarH.sec}
Although a system of $N$ protons interacting with $N$ electrons via the
Coulomb potential seems a clear
 problem in quantum statistical mechanics, it turns out to be very complex,
given the possibility of bound states of
electrons and ions, phase transitions etc. Initially, prior to first-principles
methods, various phase-transitions involving normal electronic
states and even superconducting states had been
 conjectured~\cite{Norman68,Ashcroft68}. Today, we have rigorous tools
 for studying partially or fully ionized
hydrogen plasmas. However, even
the very concept of ``the degree of ionization''~\cite{Murillo13}
 has been challenged, leading to insightful discussions
 of the topic~\cite{Gawne24,Sharma25}.  

Bonitz et al.~\cite{BonitzPOP24}
state that:\\
 ``the degree of ionization and
the fractions of atoms and molecules are not physical observables.
They cannot be rigorously computed, even by a first principles 
simulation. Any result depends on the used criterion,\ldots,
even though the sensitivity to the chosen criterion can be verified and
minimized".\\

More  specifically, such computations cannot even distinguish and
 differentiate the physical species
H, H$^-$, (H$_2)^+$, etc., that are formed through bound-electron effects
that manifest in partially ionized plasmas. The
computed $g_{\rm HH}(r)$, i.e., the proton-proton PDF $g_{\rm pp}(r)$,
 obtained from such methods, for such partially
 ionized plasmas may contain a featureless broad peak, or
usually an extra ``hump'' prior to the main peak, with no
species differentiation.

Other authors have also claimed that there is ``no quantum operator''
 corresponding to the ``degree of ionization'' in disputing the use
 of $\bar{Z}$ in WDM calculations~\cite{PironBlenski11,Pain2023,SXHu16}.
They use the temperature $T$ although
it too has no quantum operator in quantum mechanics.
 In the study of WDM, we use quantum {\it statistical} mechanics, with
 a classical heat bath attached to the system rather than
 a purely quantum mechanical system. Hence,
quantities such as the temperature $T$, the chemical potential $\mu$,
and the mean ionization $\bar{Z}$ can be introduced as Lagrange multipliers
of the theory~\cite{DWP82}. They can also be introduced into
 numerical simulations by introducing thermostats (for $T$), 
numberstats (for $\mu$),  as well as chargestats (for $\bar{Z}$).
While the use of thermostats is well developed, the other types of
controls for the number of particles or charge neutrality have
not been taken up extensively. This is partly why quantities
 like $\bar{Z}$ remain currently unacknowledged in
 first-principles simulations. 
 
Furthermore, as discussed in Ref.~\cite{CDWBe25} and in the Appendix,
a quantum operator whose mean value  gives us
a precise value of $\bar{Z}$ via the Friedel sumrule can be constructed.
The mean charge $\bar{Z}$ arises
naturally within the neutral-pseudo-atom model of DFT.
There is no difficulty in unambiguously defining $\bar{Z}$ for a strict Lorentz plasma with
no e-e and i-i interactions. The one-electron states and their
occupations are hydrogenic and provide the same $\bar{Z}$ either from the
occupation numbers or from the Friedel sumrule. When interactions are present,
there are no one-electron states, as they become Dyson quasi-particle states~\cite{PDW-Dyson84} and naive definitions of $\bar{Z}$ fail.
DFT switches off the e-e and i-i interactions by constructing fictitious
Kohn-Sham states (of a modified Lorentz plasma)
 that provide the total density and the total free energy correctly. 

The PIMC method uses Feynman trajectories but do not
attempt to distinguish what may be termed bound and free electrons paths
at any stage of its calculation.
In effect {\it it has set up no machinery to}
  identify and distinguish
species like  H, (H$_2)^+$, H$_2$, H$^-$ that exist in a partially
ionized H-plasma. That does not mean that these species have no
``physical existence" but it shows the need to extend PIMC to capture
more complex ``emergent'' statistical quantities. In fact, a broadened unresolved
``bump" appears in the $g_{\rm HH}(r)$ calculated  by current PIMC  methods. 

 In reality, the PIMC, QMD or NPA calculations,
 all of which begin with an input of electrons and nuclei, should be
 regarded as a sequence of successive  calculations that can be
used to analyze the speciation in hydrogen and other partially ionized plasmas in
 an appropriate manner, as the primary plasma, i.e., electrons and nuclei,
respond to changes in different energy and length scales. New structures
appear as new poles of the $S$-matrix (expressed in terms
of particle trajectories), and in appropriate response functions
 that can in principle be calculated within PIMC., if these emergent entities
 are to be made manifest.

Even at the static level, PIMC and QMC methods can calculate $S_{ee}(k)$
 as well as $S_{ii}(k)$.
Experimentally, while $S_{ii}(k)$ can
be obtained from X-ray or neutron diffraction, $S_{ee}(k)$ is obtained from
XRTS measurements. Then, the compressibility sumrule and charge neutrality
together lead to the following relations, enabling an unambiguous experimental
 or theoretical estimate of $\bar{Z}$.
\begin{eqnarray}
\label{comp-sumrule.eqn}
S_{NN}(k)&=&S_{ee}(k)+S_{ii}(k)+2S_{ei}(k)\\
\label{comp.eqn}
S_{NN}(k\to0)&=&(\bar{\rho}+\bar{n})T\xi\\
\label{chg-neut.eqn}
S_{ee}(k\to0)&=&2\bar{Z}S_{ei}(k\to0)-\bar{Z}^2S_{ii}(k\to0)
\end{eqnarray}
The equation~\ref{chg-neut.eqn} follows from the charge-neutrality requirement
that $\bar{n}=\bar{Z}\bar{\rho}$ and that the total charge-charge structure factor
reduces to zero as $k \to 0$. Furthermore,
When $\bar{n}g_{ei}(r)$ is identified with the free-electron
charge at an ion, these relations can be
further simplified to give
direct access to $\bar{Z}$. The free-electron
pile-up around an in the NPA is such that:
\begin{eqnarray}
\label{nf.eqn}
n_f(r)&=&\Delta n_f(r)+\bar{n}=\bar{n}g_{ei}(r)\\
\label{sei.eqn}
S_{ei}(k=0)&=&\int\left[g_{ei}(r)-1\right]4\pi r^2 dr \\
\label{barz.eqn}
\bar{Z}&=&S_{ee}(k\to0)/S_{ii}(k\to0).
\end{eqnarray}
In Eq.~\ref{comp.eqn}, $\xi$ is the isothermal compressibility of the system of
electrons and ions. The recovery of these $k \to 0$ limits of $S_{ab}(k)$ 
in a simulation or via a physical model like the NPA establishes the thermodynamic
consistency of the calculation since $\xi$ is
directly obtained from the EOS. In fact, the MHNC equation uses this relation
to check its treatment of bridge-diagram contributions to the potential
of mean force in classical-fluid theory~\cite{LFA83}. 

 PIMC and QMC, being limited by the
 simulation box size, cannot cheaply
access the $k\to 0$ limit of any of the structure factors; but
this is not an existential objection against $\bar{Z}$.
 The compressibility sumrule given above
is not used in NPA calculations to determine $\bar{Z}$ because $S_{ee}(k)$ is not 
directly available from an NPA calculation.    

\section{Metastable states and plasma-phase transitions in hydrogen}
Another much-discussed feature of partially ionized plasmas is the possible existence
of metastable states within regions of
 liquid-liquid (plasma-plasma)
 phase transitions~\cite{Ebeling85,SaumonChab89,Weir96,Margo96, eos95,cdwSi20,cdw-carb22,CDW-Pool25}.
 Bonitz et al.~\cite{BonitzPOP24}, in their review of hydrogen conclude that there is still no
hard evidence for a plasma-phase transition (PPT) in spite of many claims for it.
 Many papers that propose a PPT
use intuitive physical  models. Margo et al.,~\cite{Margo96} used PIMC-methods
and proposed in 1996 a PPT at $\sim$ 1 eV and $r_s\sim$ 2.2. This has not been confirmed by
subsequent studies. However, 
Filinov et al.~\cite{Filinov2001,Filinov2003} used PIMC methods, simulating
 50 electrons and 50 ions, and found  that at $T\sim$ 10 000 K, in the
density range of 0.1 g/cc...1.5 g/cc, strong density fluctuations and
the formation of metallic clusters (i.e., presumably regions predominantly with 
$\bar{Z}$ = 1) occurred, 
but without clear convergence of the simulations.
Here we study the 0.1 g/cc isochore and provide comprehensive
evidence for two competing plasma states, using  DFT methods.

Some of proposed PPTs involve changes in short-ranged order~\cite{NormanSait17}.
Other PPTs may not show significant changes in short-ranged order as reflected in the
PDFs. Instead, they involve the  cooperative action of a large numbers of atoms and
distant neighbours in the tails (large-$r$) of PDFS. Such transitions,
 (e.g., Martensitic transitions) are beyond the reach of simulations that employ a few hundred
nuclei. Formation of such metastable phases may occur without any
 change in $\bar{Z}$~\cite{cdwSi20,cdw-carb22,CDW-Pool25},
or they may have two different average ionizations
at a given density and temperature. In the latter case, 
 $\bar{Z}$ becomes a quick label for indicating such phases. For
instance, hydrogen at a given density and temperature may exist as a fully
ionized metastable phase with $\bar{Z} = 1$, or with a value less than unity.
Weather they ``co-exist'' with phase separation, or with the formation of
spinodal phases etc., are topics beyond the scope of the present study.

\begin{figure}[t]                    
\includegraphics[width=0.96\columnwidth]{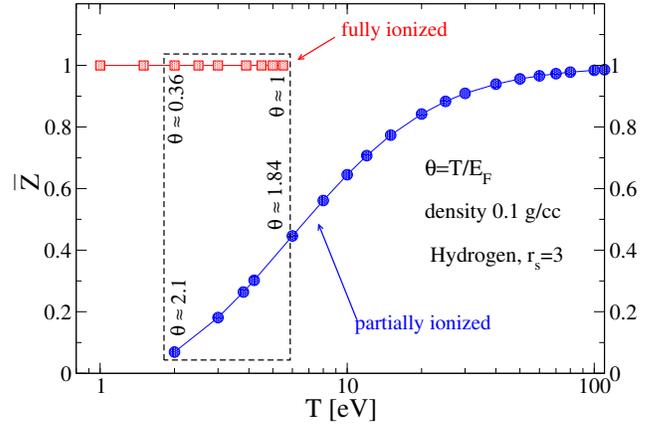}
 \caption{\label{H-partial.fig}(online color) The competition
and possible co-existence of
a partially ionized phase with $\bar{Z} < 1$, with a fully ionized
phase with $\bar{Z} = 1$ for the hydrogen isochore,
$\bar{\rho} = 0.1$ g/cc, i.e., $r_{\rm ws} \simeq 3$.} 
\end{figure}

Moldabekov et al.~\cite{MoldHyd25} have studied hydrogen at $T$ = 4.8 eV 
and density 0.08 g/cc ($r_{\rm ws} = 3.23$). Hydrogen plasmas within that range of
 densities are typical of those that have been proposed (see Fig.~4 of the review
by Bonitz et al.,\cite{BonitzPOP24}) to have PPTs due to 
competing metastable states, with
 $\bar{Z} = 1$ or $\bar{Z}<1$. In Fig.~\ref{H-partial.fig} we display the behaviour
of $\bar{Z}$ for a hydrogen plasma of density 0.10 g/cc ($r_{\rm ws} = 3)$ over  a
wide range of temperatures to illustrate the competition between a fully
ionized phase and a partially ionized phase.
The region indicated by a box, within the temperature region 2 eV to 6 eV
is characterized by two possible plasma phases that we could reach by
starting our NPA calculations from either (a) a fully ionized trial solution
or (b) from a partially ionized initial starting point. This behaviour is
found in a variety of H-plasmas for $r_{\rm ws} \sim 3$, possibly including the plasma
with $\bar{\rho}$ = 0.08 g/cc studied in Ref.\cite{MoldHyd25} using PIMC. 

\begin{figure}[t]                    
\includegraphics[width=0.96\columnwidth]{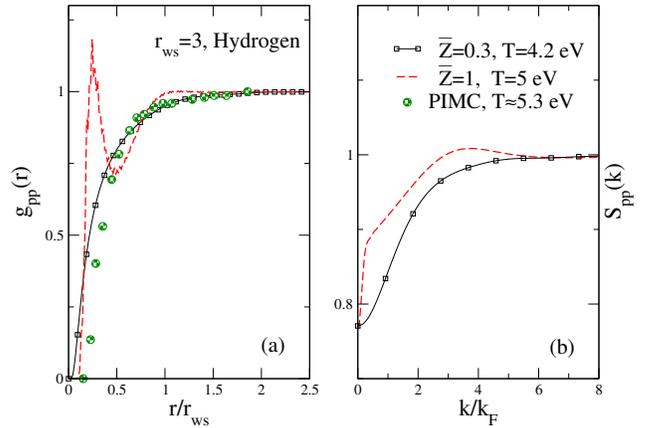}
 \caption{\label{gr-partial.fig}(online color) The proton-proton pair distribution function
$g_{\rm pp}(r)$ and the corresponding $S_{\rm pp}(k)$ for a fully-ionized
plasma ($\bar{Z} = 1$), and for a partially ionized plasma ($\bar{Z} = 0.3$)
from within the metastable region, for the density
$\bar{\rho} = 0.1$ g/cc, i.e., $r_{ws}\simeq3$. The PIMC $g(r)$ is from
Ref.~\cite{BonitzPOP24}, Fig.~18, where the conditions $r_s$ = 3, $T$ =  62500K
are specified, with $\bar{Z}$ unspecified.}
\end{figure}

In Fig.~\ref{gr-partial.fig}, we display the proton-proton PDF, viz., $g_{\rm pp}(r)$
in panel (a), and the corresponding structure factor $S_{\rm pp}(k)$ in panel (b) for
two H-plasmas of density 0.1 g/cc. We also display the closest available 
(in terms of density and temperature) restricted PIMC simulation of $g(r)$, extracted
 from Fig.~18 of Bonitz et al.~\cite{BonitzPOP24}.
The $g(r)$ with a sharp peak is on the fully-ionized branch,
while the other is on the partially ionized branch. The ``physical reason" for
the existence of these two possibilities becomes clear from their PDFs.
\begin{figure}[t]                    
\includegraphics[width=0.96\columnwidth]{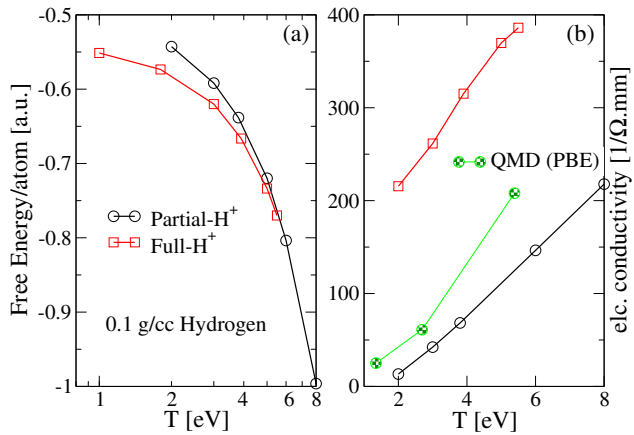}
 \caption{\label{F-sig.fig}(online color) (a) The Helmholtz free energy $F$ per atom (Hartrees)
for the partially-ionized plasma (labeled partial-H$^+$), compared with the
fully ionized plasma (labeled full-H$^+$). (b) The electrical conductivities
of the partially-ionized and fully-ionized plasmas, both of density 0.1 g/cc 
are displayed, calculated using a T-matrix approach
suitable for strong scattering. The QMD values of $\sigma$ are from the
calculation by Holst at al.~\cite{Holst11} using the $T=0$ PBE-XC functional.}
\end{figure}

 The $g(r)$ of the fully-ionized system  has a sharp peak near $r$ = 0.75
 atomic units, and also a broad peak near 
$r = 3$ atomic units. The convergence in this region is difficult. Although
the ion-ion potential is repulsive, by placing a nearest-neighbour
 proton at $r < 1$ a.u., the system prevents the formation of bound electrons
in atomic or molecular states; but it  gains stability by increased XC-energy,
 in holding a  high free-electron density, with the plasma remaining 
in the fully ionized state at $r_s = 3$ and $r_{\rm ws}$ = 3. The alternative possibility is to
allow atomic and quasi-molecular forms of hydrogen to form, gaining stability
by reducing the ion-ion electrostatic repulsion, while sacrificing the XC-energy
when the free-electron density decreases, with $r_s$ dropping
from 3 to 4.48 while $r_{\rm ws}$ remains 3.

To obtain a converged PDF on the partially ionized branch at 5 eV itself proved
difficult. However, the results at $T$ = 4.2 eV, $\bar{Z}$ = 0.3
 are presented (curve marked with  squares as data points) as being typical
of this regime.
 Here, although  $r_{\rm ws}$ = 3, $r_s$ is 4.48 as 70\% of the electrons are
 in  bounds states. The interactions among the bound forms (quasi-molecules)
 are weak, and hence the $g_{\rm pp}(r)$ is seen to be rather featureless, although
 screening is also weak at  $r_s$ = 4.48. As we could not find a PIMC
or QMD calculation at the temperatures of interest, we display the
case $T\sim$ 5.3 for which PIMC calculations are displayed in Fig.18 of 
Ref.~\cite{BonitzPOP24}. While the PIMC $g(r)$ falls on the partially-ionized
branch for larger $r$, there is disagreement in the small-$r$ region. The
origin of this difference is not clear at present.

The competition between the two hydrogen phases, namely, full-H$^+$ and partial-H$^+$,
is best quantified by comparing their  Helmholtz free energies per atom. 
In panel (a) of fig.~\ref{F-sig.fig} we display the
 Helmholtz Free energy $F(r_{\rm ws},r_s, \theta)$
calculated as in Eq.~\ref{Ftot.eqn}. We have  omitted
 the ideal classical gas free energy in the plot as it is the same
for both phases. 
\begin{equation}
\label{Ftot.eqn}
F(r_{\rm ws},r_s,\theta)=F_{\rm uef}(r_s,\theta)+F_{emb}(r_{\rm ws},r_s,\theta)+F_{12}+F_{ideal}
\end{equation}
The first term in $F$ is the contribution from the UEF contained in each phase. This
depends on $\bar{Z}$ which determines $r_s$. The second term, $F_{emb}$ is the energy of embedding
a proton in the electron fluid. This process involves (i) the creation of a bound core of electrons, (ii)
a modification of the continuum density of states which are occupied by the electron-charge pileup
$\Delta n_f$ of Eq.~\ref{nf.eqn}, and (iii) setting up the particle-density profiles $n(r),\rho(r)$ around
the nucleus, creating  pseudo-ions of charge $\bar{Z}$. The term $F_{12}(r_{\rm ws},r_s,\theta)$
represents the contributions from ion-ion interactions. The details of how these terms could be calculated
relatively straight-forwardly are given in Refs.~\cite{Pe-Be,eos95}.

The comparison of the Helmholtz free energies
of the two phases shows that there is a region within $T$ = 4 to 6 eV where the free energies of the 
two phases become comparable. Furthermore, $\bar{Z}$ plays the role of a miscibility parameter in this
competing system of two phases. When we consider a partially ionized H-plasma, e.g., at $T$ = 4.2 eV with
$\bar{Z}$ = 0.302, it implies that the various  charge-neutral species of hydrogen found
 in the partially ionized plasma will tolerate a mixing of  ionized components 
(e.g., H$^+$, (H$_2$)$^+$ etc.) up to about $\sim$ 30\%. 
Since $\bar{Z}$ has been calculated by a minimization of the Helmholtz free energy 
(via the energy-minimum  principle associated with DFT), any increase or decrease in $\bar{Z}$
(by dissolving or precipitating protons) is energetically unfavorable and triggers phase separation.

\begin{figure}[t]                    
\includegraphics[width=0.96\columnwidth]{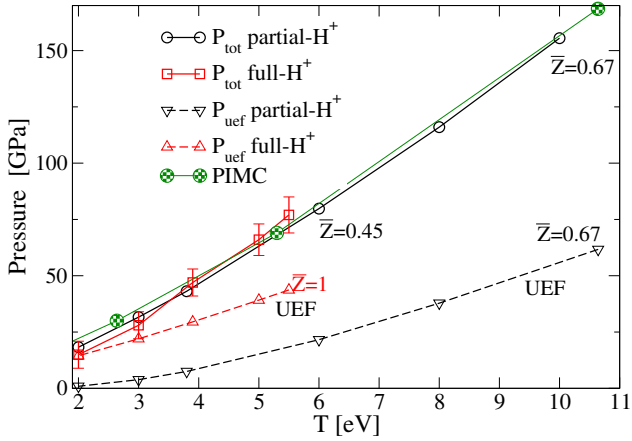}
 \caption{\label{Ptot.fig}(online color)  The total pressures along the 0.1 g/cc isochores of the two
phases, calculated from the Helmholtz free energy $F$ are displayed. The
UEF contributions are also displayed. The PIMC data are from Militzer et al.~\cite{MilitzerCataldo21}.}
\end{figure}

Given the total free energy, all other EOS quantities can be evaluated. In Fig.~\ref{Ptot.fig} we
display the pressure for the fully- and partially- ionized states along the 0.1 g/cc isochore. We also
display the sum of the kinetic pressure $p_0$ and the finite-$T$ XC-pressure $p_{xc}$ as $P_{\rm uef}$
 for the two cases.
It is seen that $P_{\rm uef}$ is a significant component of the total pressure, especially for the
fully-ionized phase. The classical ideal gas pressure is the same for both phases, and is included
in $P_{tot}$ although the corresponding $F_{ideal}$ was not included in the free-energy plot.
PIMC calculations for the pressure at 0.1 g/cc are available form Militzer et al.,~\cite{MilitzerCataldo21},
and from Table VII and Fig.14 of Ref.~\cite{BonitzPOP24}. Since these authors do not calculate
a value for $\bar{Z}$, they refer to the density of the system using $r_s$ = 3, whereas we specify
the system with $r_{\rm ws}$ = 3, while the value of $r_s$ depends on $\bar{Z}(T)$.
Our calculated pressures (along the two metastable branches) are very close to each
other within numerical uncertainties. The pressure of the partially ionized branch, which is the
 only one that survives for $T >$ 6 eV agrees closely with the PIMC pressure data. 

Fig.~\ref{F-sig.fig}(b) displays the electrical conductivity $\sigma$ of
 the partially-ionized phase, compared
 with  the fully ionized phase, calculated using finite-$T$ XC, i,e., PDW-XC, and 
a T-matrix to include strong scattering.
 The fully ionized phase has a high conductivity.
The QMD calculations of $\sigma$ by Holst et al.,~\cite{Holst11} use a $T$ = 0 PBE-XC functional. QMD
 does not distinguish two possible plasma phases. Given the well-known sensitivity of QMD-$\sigma$
 to the XC-functional, the agreement is satisfactory.  
 
The picture of the hydrogen plasma used in our calculations is that of many pseudo-ions of charge
$\bar{Z}$ (carrying their ionic- and electronic- screening envelops) interacting with one another. If we
consider the plasma at $T$ = 6 eV, $\bar{\rho}$ = 0.1 g/cc, these pseudo-ions have a charge of 0.446.
However, this does not mean that {\it in reality}, there are such pseudo-ions of charge 0.446 in the
plasma at any instant. What it means is that
the long-time thermodynamic average of various types of ions and neutrals
that exist in the plasma can be represented
as a statistically averaged object with a charge of 0.446. Such a statistical model gives a pressure
(and other properties such as X-ray structure data) in very close agreement with that from PIMC
 where the averaging is carried over particle trajectories. One may ask if a
 proper book-keeping of electron trajectories
may perhaps be used to construct these pseudo-ions from the trajectories
to recover the DFT one-body picture.

 Since the DFT used in
the NPA reduces the many-body problem to effective non-interacting
one body objects (electrons, pseudo-ions),
it does not construct have objects like (H$_2$)$^+$ although their effect on the densities and free energies are fully accounted for. In contrast, the PIMC has not made such a reduction of the many-body problem and it should, in principle,
be able to develop methods of identifying complex multi-center objects such as  (H$_2$)$^+$, H$_2$ et cetera.
  
\section{Classical Map calculations for the UEF and H-plasmas}
\label{clm.sec}
Although the classical-map (Cmap) method is not as well grounded as the
NPA method for general WDMs, the CMap works well for fully ionized plasmas.
 Furthermore, it provides a very accurate treatment of the UEF and was used to construct
a sophisticated e-e XC functional~\cite{prl1} prior to the advent of QMC and PIMC-based
XC functionals. The method has been successfully applied to calculating
XC-energies and pair-distribution functions of
 reduced dimensional systems that arise in the physics of 
nano-structures~\cite{prl2,prl3,Jost05,Totsuji09,lfc1-dw19}.
In fact, since the current NPA code uses the finite-$T$ PDW-XC functional, here
we  establish that it is adequate for the calculations presented
here, using comparisons with the functional of Groth et al.~\cite{GDS17}. 

\begin{figure}[t]                    
\includegraphics[width=0.96\columnwidth]{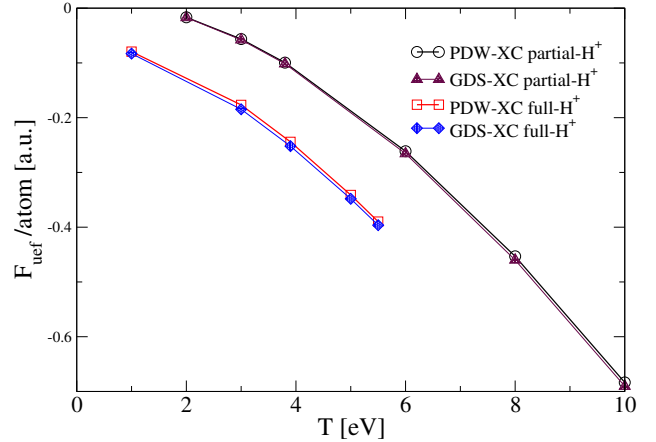}
 \caption{\label{Fuef.fig}(online color) The
UEF components of the total Helmholtz free energy $F$ per atom (Hartrees)
for the partially-ionized plasma (labeled partial-H$^+$), and for the
fully ionized plasma (labeled fully-H$^+$) at 0.1 g/cc, calculated using the
classical map (i.e., PDW-XC)~\cite{PDWXC}, and via the parametrization
 given by Groth et al. (GDS-XC)~\cite{GDS17} are displayed.}
\end{figure}

A major contribution to the Helmholtz free energy $F$ given in  Eq.~\ref{Ftot.eqn} is the
UEF component $F_{\rm uef}(r_s,\theta)$. We calculate $F_{\rm uef}$ using the PDW-XC fit used
in the NPA, as well as with the fit formula for $F_{\rm uef}$  given by Groth et al. (GDS-XC) for
the two phases, partial-H$^+$ and full-H$^+$. The
results are presented in Fig.~\ref{Fuef.fig}, and cover a regime of $r_s$ and $T$ such that
$1< r_s <8$, and 1 eV $< T < 10$ eV. As the stabilization of the fully-ionized H$^+$ plasma
against the partially-ionized plasma is largely controlled by the free energy of the UEF, the good
agreement between $F_{\rm uef}$ calculated in the NPA using PDW-XC, with that of GDS-XC shows 
(i) that the Cmap provides a reliable XC functional for these WDM calculations, and (ii) that the
identification of two competing plasma states reported here is not likely to be
 modified by the use of alternative e-e XC-functionals.

The Cmap for the UEF requires only
the electron parameter $r_s$, the spin polarization
$\zeta$  and the physical temperature as inputs. The spin polarization
will be taken as zero, as WDM studies have currently
addressed mainly the case $\zeta = 0$. However, the capacity to do
spin-polarized calculations is important since many giant planets
and other astrophysical objects have strong  magnetic fields.

\subsection{Calculations using the Classical Map for the UEF}
\label{clmUEF.sec}
The importance of the e-e structure factor $S_{ee}(k)$ has been emphasized
in recent work~\cite{MoldHyd25}. We also noted  the relevance of $S_{ee}(k\to0)$ in regard
to the concept of $\bar{Z}$ (see Eq.~\ref{barz.eqn}) and the compressibility
sumrule .

There have been many calculations of the structure factors
 $S_{e,e}(k,\zeta)$ and PDFs of the uniform electron fluid at $T$ = 0, using
mainly fixed-node Monte Carlo, and diffusion Monte-Carlo methods. Comparisons
of results from the CMap for $T$ = 0, both for $\zeta = 0$ and $\zeta \ne  0$,
were given in previous publications~\cite{prl1}. 
In this study we limit ourselves to more recent finite-$T$ UEF calculations.
In this context we select the following densities $\bar{n}$
and temperatures $T$, as PIMC H-plasma calculations are also
available  for most of them. We use $\theta = T/E_F$ in the following.

1. $r_s = 2$, i.e., nominally $\bar{n}$ = 0.33 g/cc at $T$ = 4.8 eV.
 In
Ref.~\cite{MoldHyd25}, $\theta$ = 1, 0.5, and 0.25 are equated with
$T$ = 12.5 eV, 6.27 eV and 3.13 eV. This is consistent with an assumed
ionization of $\bar{Z} = 1$ in hydrogen. NPA calculations at $r_s = 2,
T$ = 12.5 eV, 6.25 eV, and 3.13 eV confirmed that they can be
treated as fully ionized plasmas with $\bar{Z} = 1$. No co-existence
or competition of partially ionized states was encountered.

However, Ref.~\cite{MoldHyd25}
also uses a 'chemical model' with $\bar{Z}$ = 0.72, 0.64, and 0.61. The
case $\bar{Z}$ = 0.72  should correspond to $\theta$ = 1.26, $r_s$ = 2.247,
while $\bar{Z}$ = 0.64 implies  $\theta$ = 0.683, $r_s$ = 2.337; 
and $\bar{Z}$ = 0.51 implies that $\theta$ = 0.352, $r_s$ = 2.375.
The authors have not indicated how these $\bar{Z}$ values
have been obtained. We do not examine these cases any further in this
study. 

In the case the UEF, we do not concern ourselves with $\bar{Z}$,
and we simply use $r_s$ = 2, $\theta$ = 1, 0.5 and 0.25 for comparisons with the
results given in the literature. In the case of H-plasmas, a first-principles
determination of $\bar{Z}$ becomes necessary. Ref.~\cite{MoldHyd25} has not
reported PIMC calculations for $\theta$ = 0.5, 0.25 as PIMC is not viable
at low temperatures. They have provided results based on a
{\it new ansatz} combining time-dependent DFT results for the dynamic structure
factor $S_ee(q,\omega)$ with static DFT results for the density response. 
 
2.  $r_s = 3.23$, i.e., nominally $\bar{n}$ = 0.08 g/cc at $ T/E_F$ = 1,
and given as 4.8 eV in Ref.~\cite{MoldHyd25} imply a value of $\bar{Z}$ = 1.
 Furthermore,  Ref.~\cite{MoldHyd25} refers to a `chemical model' (CM),
 where a mean ionization of $\bar{Z}$ = 0.57 is used. If $\bar{Z}$ = 1 were
assumed, then setting  $\theta$ = 1 is consistent. However, the value
 of $\bar{Z}$ for this hydrogen plasma
(i.e., 0.08 g/cc at 4.8 eV), obtained  using the Friedel sum rule implemented
via the NPA code gives $\bar{Z}$ = 0.365, significantly lower than the
chemical models used by Moldabekov et al.~\cite{MoldHyd25}. If the NPA
value of $\bar{Z}$ is adopted then, $r_s$ = 4.519, while $\theta$ = 1.956.
Hence we are unable to make a proper comparison with the results given
in Fig. 1  of the study by Moldabekov et al~\cite{MoldHyd25}.

\begin{figure}[t]                    
\includegraphics[width=0.85\columnwidth,  angle=-90]{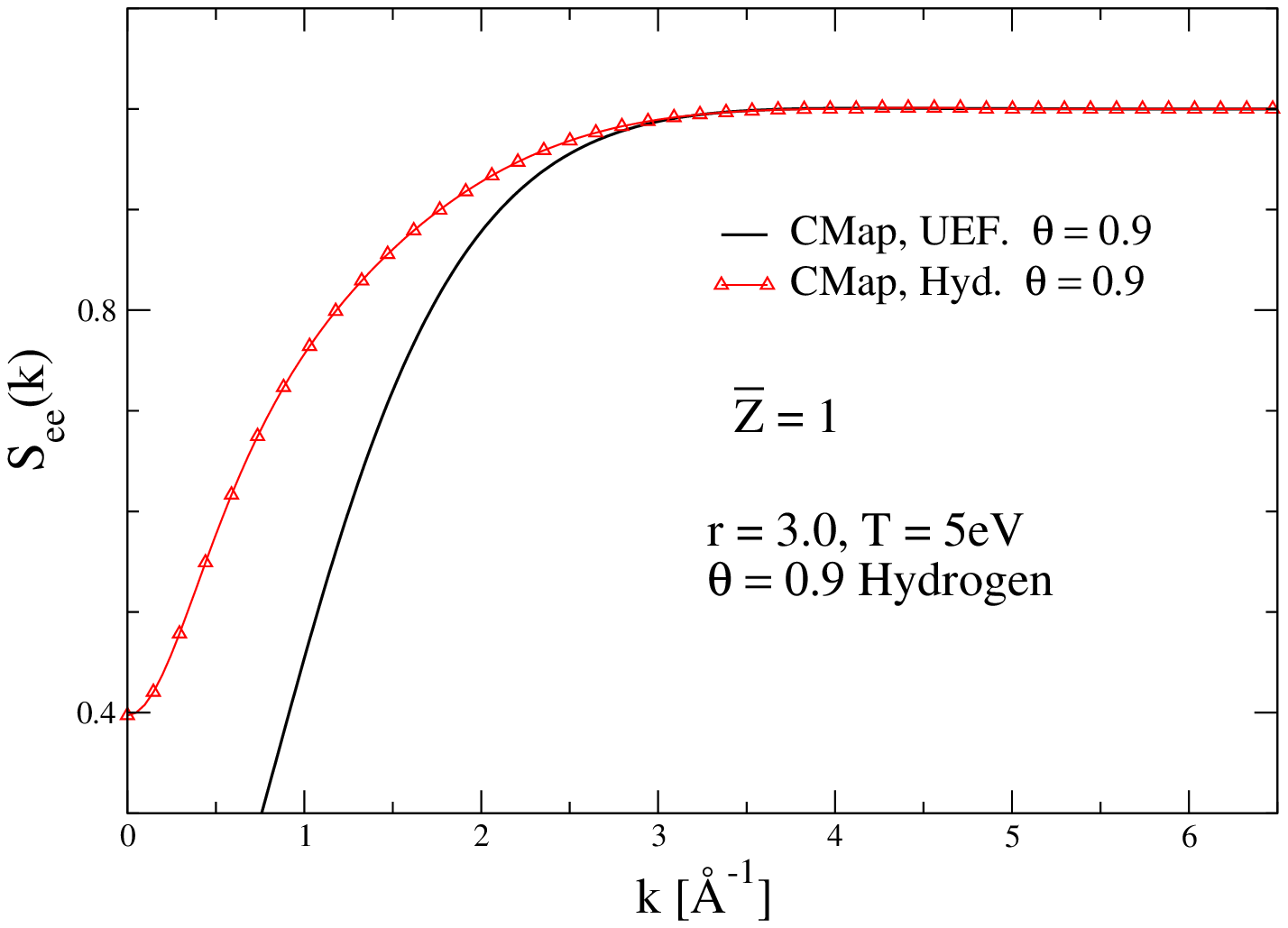}
 \caption{\label{rs3tef0p9.fig}(online color) The e-e structure factor $S_{ee}(k)$
at $r_s = 3, \theta = 0.897$ for the uniform electron fluid (UEF),
 and for fully ionized
hydrogen calculated using the Classical Map (CMap) method.}
\end{figure}

However, as seen from Fig.~\ref{H-partial.fig} for hydrogen at 0.1 g/cc,
 a density of 0.08 g/cc at 4.8 eV is a good candidate for being in a
fully ionized state {\it and also} in a partially ionized state. Unfortunately,
even on starting with fully ionized trial inputs, we could not generate
Kohn-Sham NPA solutions at 0.08 g/cc and 4.8 eV that were fully ionized. 
As a somewhat close approximation to the density (0.08 g/cc)
studied by Moldabekov et al., in Fig.~\ref{rs3tef0p9.fig} we display
 the $S_{ee}(k)$ for fully ionized hydrogen, and the corresponding UEF,
 at $r_s$ = 3, $\bar{\rho}$ = 0.1 g/cc
and at 5.0 eV.

\section{Classical Map and NPA calculations for H-plasmas}
Moldabekov et al.~\cite{MoldHyd25} provide PIMC calculations for $S_{ee}(k)$
for hydrogen plasmas at the density 0.33 g/cc, and at $\theta$ = 1, 0.5 and 0.25.
We have used the nomenclature ``uniform electron fluid'' (UEF) for quantum electron 
systems with a uniform non-responding neutralizing positive background. The 
name ``uniform electron gas'' is also used by many authors, most appropriately if $r_s < 1$.
As already noted, it has been shown~\cite{prl1} with good accuracy that the quantum electron fluid
at a physical temperature $T$ is statistically equivalent to a classical
Coulomb fluid at the temperature $T_{cf}$, Eq.~\ref{tcf.eqn}. Now that results
for $F_{xc}(T)$ for the UEF are readily available, the ansatz used in 
Eq.~\ref{tcf.eqn} is not necessary and $T_{cf}$ can indeed be calculated
 directly, but choosing
$T_{cf}$ to recover the known $F_{xc}(r_s,T)$ from a classical-fluid
calculation. However, in this study we
retain the original ansatz whose accuracy was checked in Fig.~\ref{Fuef.fig}.

\subsection{Results for $r_s = 2$, $\theta = 1$}
\begin{figure}[t]                    
\includegraphics[width=0.85\columnwidth, angle=-90]{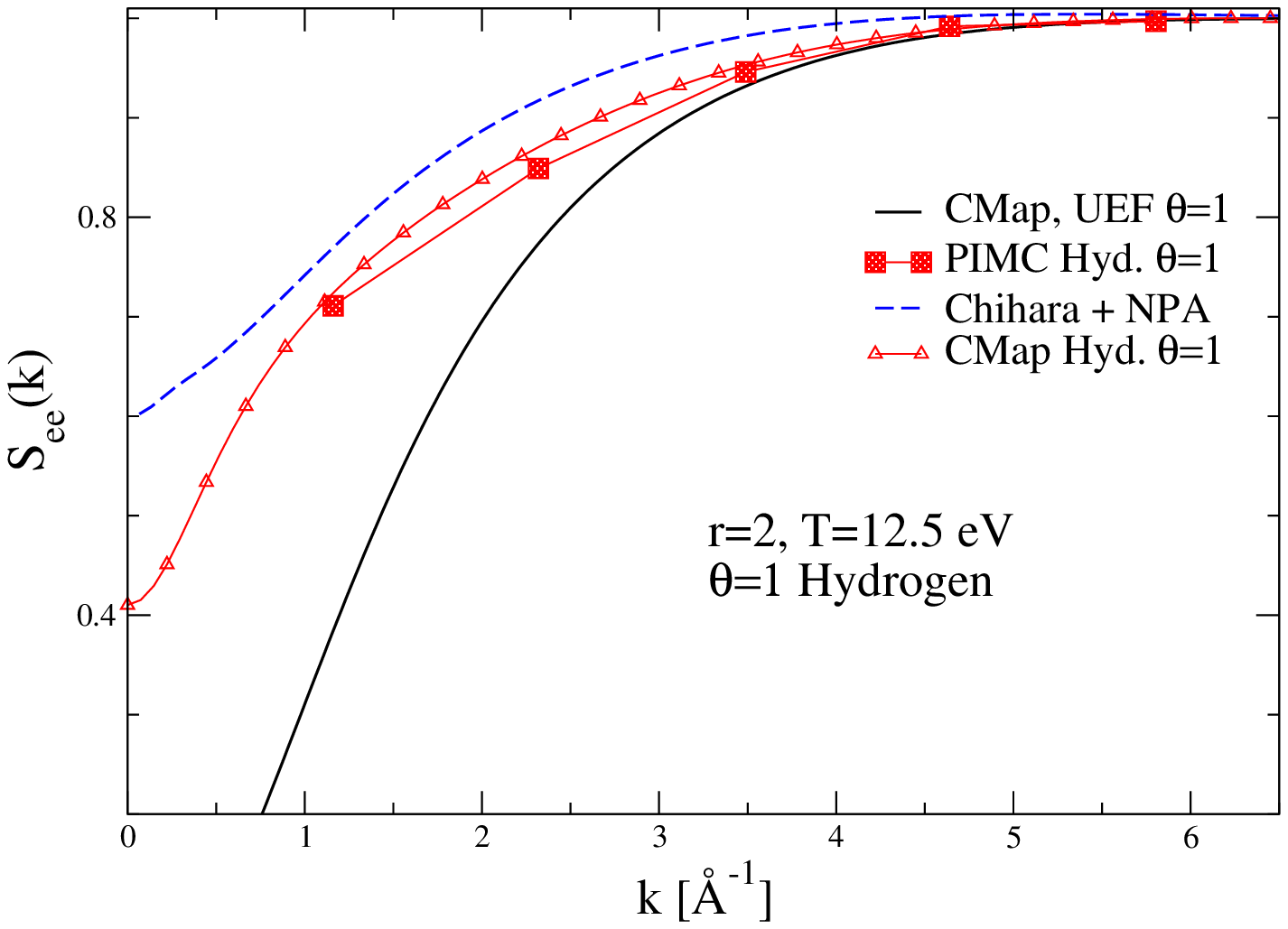}
 \caption{\label{0p33gTEF1.fig}(online color) The e-e structure factor $S_{ee}(k)$
at $r_s$ = 2, $\theta$ = 1 for the uniform electron fluid (UEF), and for fully ionized
hydrogen. Solid black line is $S_{ee}(UEF,k)$ for the UEF calculated using the
 CMap approach. The corresponding result for fully ionized
hydrogen (red line with triangles) is to be compared with
the PIMC results (filled red squares) reproduced from  Ref.~\cite{MoldHyd25}.
 The dashed blue line is $S_{ee}(k)$ constructed according to Chihara's
 prescription, with the relevant quantities calculated using the NPA method.}  
\end{figure}

Figure~\ref{0p33gTEF1.fig} is essentially similar to Fig.~4 of Ref.~\cite{MoldHyd25},
and treats the case $\theta = 1$, $\bar{\rho}$ = 0.33 g/cc, at 12.5 eV which
 corresponds to $r_s = 2$ since the system is fully ionized. 

An approximate form of $S_{ee}$ proposed by Chihara~\cite{Chihara2000}
 can also be calculated using
 the  so-called  ``ion feature" $I(k)$  of XRTS theory, using the
bound electron-density $n_b(k)$ and the
screening density $n_f(k)$ that is formed around each ion of nuclear charge
$Z_n$. In the present case there are no bound electrons,
with $\bar{Z} = 1$ and $n_b(k)=0$.
\begin{eqnarray}
\label{SeeChihara.eqn}
I(k)=\{n_b(k)+n_f(k)\}^2S_{ii}(k)\\  
S_{ee}(k)=S_{ee}(UEF,k)+I(k)/Z_n
\end{eqnarray}
Chihara's approximation does  not satisfy the compressibility sumrule for
a two component fluid unless the compressibility of the UEF is negligible.
We have calculated the relevant quantities, viz., $n_f(k), S_{ii}(k)$ using the NPA
and constructed $S_{ee}$ according to Eq.~\ref{SeeChihara.eqn}. This is displayed
(dashed blue line) in Fig.~\ref{0p33gTEF1.fig}. We find that the Chihara approach
 is only moderately successful here, unlike with more dense plasmas, e.g., highly
compressed Be studied in Ref.~\cite{CDWBe25}. In contrast, the results of the
CMap approach  seem to be in agreement. Furthermore, the Cmap, if applicable, provides
an inexpensive approach to extending the limited structure data from
PIMC and QMD towards the $k \to 0$ limit.

\section{Conclusion}
Our study of hydrogen plasmas presented here has demonstrated the existence
of competing  states of ionization in partially ionized hydrogen plasmas in the
liquid state.
For instance, for  $r_{\rm ws}\sim 3$ and for temperatures such that $0.35 < T/E_F \le 1$,
two competing energy and length scales arise wherein a high-density electron fluid
stabilized by its XC-energy competes with a partially ionized plasma with a high
embedding energy. The mean ionization $\bar{Z}$ is also interpreted to be a measure
 of the degree of miscibility of the fully ionized H-phase in the completely un-ionized
 (covalent-like) H-phase.

 We  used a first-principles method,
namely, one-atom density functional theory where both the e-e and ion-ion many-body problems
have been reduced to one-body problems. This one-body DFT approach,
 computationally realized as the NPA model,  uses two
XC-functionals (one for electrons, and another for ions), where the e-e functional
is taken in as an external input. 

We have also
used a classical-map scheme where the electron subsystem is replaced by
 a statistically equivalent classical Coulomb-type fluid whose temperature is
 selected to reproduce the correlation energy of the quantum fluid at $T$ = 0.
The exchange energy is exactly treated by the selection of the non-interacting
PDFs of the finite-$T$ UEF. This classical-map approach allows a
 convenient, numerically very fast
treatment of fully ionized  hydrogen (at higher temperatures), whereby all the structure factors
 (including the  electron-electron structure factor) can be computed.  

The further characterization of competing metastable states of ionization
in partially ionized H-plasmas (as revealed in this study) is important, both from
the point of view of basic physics, as well as in regard to applications
in astrophysics and high-energy density physics. Furthermore, the development of
theoretical tools to expose and identify the atomic and quasi-molecular species
 (such as  H, H$^-_2$, H$_2$, H$^-$ etc.) that exist in partially ionized plasmas,
but not revealed by current PIMC and QMC calculations,  is also an important task
 for future research.  

\appendix
\section{}
\subsection{The finite-$T$ Friedel sumrule and $\bar{Z}$}
\label{Friedel.sec}
Unlike ion-ion scattering which is largely repulsive, electron-ion scattering is
attractive and close interactions and multiple scattering effects are important.
The continuum wavefunctions are asymptotically like phase-shifted plane waves.
 Denoting the radial part of the free-electron-like 
Kohn-Sham functions of energy $\epsilon_k=k^2/2$
by $R_{kl}(r)$, we have the asymptotic form:
\begin{equation}
R_{kl}(r)|_{r\to\infty}\simeq \sin\left[kr-l\pi/2+\delta_l(k)\right]/kr
\end{equation}
where $\delta_l(k)$ is the phase shift. 
The phase shift is related to the scattering operator $\hat{S}_l(k)$ of the $S$-matrix
by the relation~\cite{FanoRau86}:
\begin{equation}
\hat{S}_l(k) = \exp\{2i{\tilde{\delta}_l(k)\}}
\end{equation}
The phase shift is a number, an observable that connects the
incoming  wave with the outgoing wave. The scattering operator
acting on the incoming wave transforms it to the phase shifted outgoing
wave. 

Keeping in mind that $\delta_l(k)$ is an observable
directly related to the scattering operator,
we may  define, for convenience,  a phase-shift
operator $\hat{\delta}_l(k)$ such that its meanvalue
is the $c$-number $\delta_l(k)$.
We may also define a thermal and quantum averaging of
an operator via:
\begin{equation}
\langle \hat{g}(\epsilon)\rangle =
\int_0^\infty d\epsilon\{-\frac{df(\epsilon)}{d\epsilon}\}<\hat{g}(\epsilon)>
\end{equation}
We consider the sum over the continuum states averaged over the thermal
distribution. The Friedel sumrule asserts that this sum is equal to the
charge of the scattering center, identified as $\bar{Z}$. The $T$ = 0
form was given by Friedel in 1952~\cite{Friedel52}, while the finite-$T$
version was given in 1982 in Ref.~\cite{DWP82}.
\begin{eqnarray}
\bar{Z}&=&\frac{2}{\pi T}\int_0^\infty k f(\epsilon_k)\{1-f(\epsilon_k)
\} \\
       & &\times \sum_l(2l+1)\delta_l(k)dk \\
f(\epsilon_k)&=&1/\left[1+exp\{(k^2/2-\mu^0)/T\}\right].
\end{eqnarray}
In the NPA, as we are dealing with non-interacting (Kohn-Sham) electrons, the 
non-interacting chemical potential $\mu^0$ and the unmodified density of
states is used at this stage. Furthermore, we have
given a scattering-operator type formulation as some authors have claimed that
there is ``no quantum operator" corresponding to $\bar{Z}$. The above
average is in fact a quantum {\it statistical} average involving the
thermal distribution as well.

The phase shifts are used for calculating the value of $\bar{Z}$ in each iteration
of the NPA equations until consistency is obtained. The phase shifts are also used to
calculate the modified density of continuum states that is used in evaluating the
embedding free energy appearing in Eq.~\ref{Ftot.eqn}, and in the T-matrix evaluation
of the conductivity.

\subsection{The Kohn-Sham energy levels and phase shifts of the NPA at $T$ = 10 eV, 0.1 g/cc}
As an example of the Kohn-Sham structure of a neutral-pseudo-atom
for hydrogen, we consider the case of $T$ = 10 eV, $\bar{\rho}$ = 0.1 g/cc.
In this case, the Friedel sum gives $\bar{Z}$ = 0.645.

Only one bound-electron energy level, viz.,  $1s$ exists under these conditions,
with $\epsilon_{1s}$ =0.16405 Ryds, with $f(\epsilon)$ = 0.2118. The mean
radius $<r>$ of this state is  1.8830 a.u. 

The Kohn-Sham states $|n,l,\epsilon>, |k,l,\epsilon>$ are a complete set that describes
electronic states in the field of a proton {\it and} its accompanying density distributions
 $\rho{r}, n(r)$, that together from the pseudoatom. This means, even the effects of (H$_2$)$^+$ and more complex arrangement of protons and electrons consistent with $\rho(r)$ are contained in the electron distribution $n(r)$  predicted by the above (complete set of) Kohn-Sham states. The observables $\rho(r), n(r)$ and $\bar{Z}$ are static averages. The 1$s$ energy level and its energy shfit (compared to that in atomic hydrogen) are that of a fictitious electron and need not agree with spectroscopic measurements.

\begin{figure}[t]                    
\includegraphics[width=0.96\columnwidth]{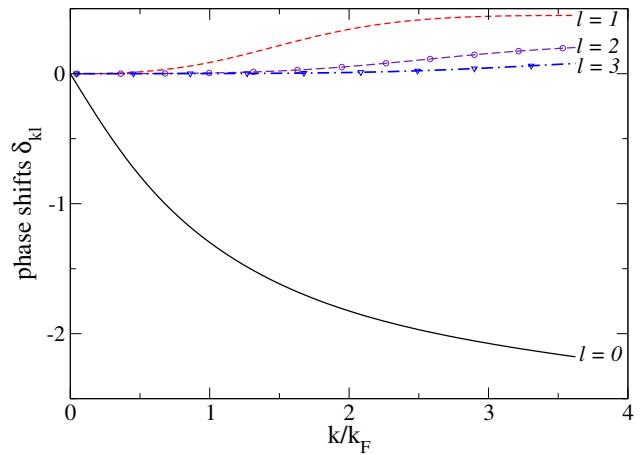}
 \caption{\label{phaseshifts.fig}(online color) The phase shifts of the $l=0, 1, 2, 3$
 angular momentum states of the free-electron wavefunctions of the neutral pseudo-atom
 of a hydrogen plasma for the case of $T$ = 10 eV and density 0.1 g/cc.}
\end{figure}

Phase shifts of angular momentum states from 0 to 3 are
presented as a function of $k$ in Fig.~\ref{phaseshifts.fig}. They become negligible for high
energy states and high angular momentum  states. A sufficient number of $k,l$ states
are used in the numerical work, and even higher-energy  contributions are
treated by Thomas-Fermi theory. These details are given in Ref.~\cite{Pe-Be,eos95}. 


\begin{thebibliography}{99}

\bibitem{Corkum07}
P. B. Corkum, {\it Attosecond science} Nature physics {\bf 3} 381 (2007).

\bibitem{Kremp99}
D. Kremp, T. Bonarth, M. Bonitz and M. Schlanges.
Phys. Rev. E {\bf 60}, 4725 (1999).

\bibitem{Ng95}    
 A. Ng, T. Ao, F. Perrot, M.W.C. Dharma-wardana, M.E. Foord,
Laser and particle beams, {\bf 23}, 527-537 (2005).

\bibitem{BonitzBk16}
M. Bonitz, {\it Quantum Kinetic Theory}, Springer (2016).

\bibitem{BonitzPOP24}
Michael Bonitz, Jan Vorberger,Mandy Bethkenhagen,
 Maximilian P. B\"{o}hme, David M. Ceperley, Alexey Filinov,
 Thomas Gawne, Frank Graziani,  Gianluca Gregori,
Paul Hamann, Stephanie B. Hansen, Markus Holzmann,
 S. X. Hu, Hanno Khelert,
Valentin V. Karasiev, Uwe Kleinschmidt, Linda Kordts,
 Christopher Makait, Burkhard Militzer,
Zhandos A. Moldabekov, Carlo Pierleoni, Martin Preising,
 Kushal Ramakrishna,
Ronald Redmer, Sebastian Schwalbe, Pontus Svensson,
 Tobias Dornhei/m.
Physics of Plasmas {\bf  31}, 110501 (2024).

\bibitem{FilinovBonitz23}
A. Filinov and M. Bonitz, 
Phys. Rev. E {\bf 108}, 055212 (2023).


\bibitem{Poole24}
H Poole, M. K. Ginnane, M. Millot, H. M. Bellenbaum, 
G. W. Collins, S. X. Hu,  D. Polsin, R. Saha, J. Topp-Mugglestone,
T. G. White, D. A. Chapman, J. R. Rygg, S. P. Regan,
and G. Gregori.
Physical Review Research {\bf 6}, 023144 (2024).
DOI: 10.1103/PhysRevResearch.6.023144

\bibitem{Drake2018}    %
 R.P. Drake, High-Energy-Density Physics: Foundation of
Inertial Fusion and Experimental Astrophysics, Graduate
Texts in Physics (Springer International Publishing,
2018).


\bibitem{Betti2016}
 R. Betti, O. A. Hurricane, Inertial-confinement fusion with lasers, 
Nature Physics {\bf 12}, 435-448 (2016).

\bibitem{GaffneyHDP18}
J.A. Gaffney, Suxing Hu, P. Arnault..E. Zurek et al.
High Energy Density Physics, Aug 2018
\url{https://doi.org/10.1016/j.hedp.2018.08.00}


\bibitem{McBride-Si-19}
E. E. McBride, A. Krygier, A. Ehnes, E. Galtier, M. Harmand, Z. Kon\^{o}pkov\'{a},
 H. J. Lee, H.-P. Liermann, B. Nagler, A. Pelka, M. R\"{o}del, A. Schropp, R. F. Smith,
 C. Spindloe, D. Swift, F. Tavella, S. Toleikis, T. Tschentscher, J. S. Wark
 and A. Higginbotham. Nature Phys. {\bf 15}, 89-94 (2019). 

\bibitem{HARB-DSF18}
L. Harbour, G. D. F\"{o}rster, M. W. C. Dharma-wardana, and Laurent J. Lewis.
Phys. Rev. E {\bf 97}, 043210 (2018).

\bibitem{eos95}
F. Perrot and M.W.C. Dharma-wardana,
Phys. Rev. E. {\bf 52}, 5352 (1995). 

\bibitem{Stanek24}
L. J. Stanek, A. Kononov, S. B. Hansen, et al.
Phys. Plasmas, {\bf 31}, 052104 (2024).


\bibitem{Gregori03}
G. Gregori, S. H. Glenzer, W. Rozmus, R. W. Lee, and O. L. Landen. 
Phys. Rev. E {\bf 67}, 026412 (2003).

\bibitem{GlenRed09}
S. H.  Glenzer and Ronald  Redmer, Rev. Mod. Phys. {\bf 81}, 1625 (2009).


\bibitem{Doppner23}
 T. D\"{o}ppner, M. Bethkenhagen, D. Kraus, P. Neumayer,
D. A. Chapman, B. Bachmann, R. A. Baggott, M. P.
B\"{o}hme, L. Divol, R. W. Falcone, L. B. Fletcher, O. L.
Landen, M. J. MacDonald, A. M. Saunders, M. Sch\"{o}rner,
P. A. Sterne, J. Vorberger, B. B. L. Witte, A. Yi, R. Redmer,
 S. H. Glenzer, and D. O. Gericke.
Nature {\bf 618}, 270-275 (2023).


\bibitem{Dornheim25}
T. Dornheim, H. M. Bellenbaum, M. Bethkenhagen, S. B. Hansen,
 M. P. B\"{o}hme, T. D\"{o}ppner, L. B. Fletcher, Th. Gawne,
 D. O. Gericke, S. Hamel,  D. Kraus, M. J. MacDonald,
Zh. A. Moldabekov, Th. R. Preston, R. Redmer, M. Sch\"{o}rner,
 S. Schwalbe, P. Tolias, and J. Vorberger.
Phys. Plasmas {\bf 32}, 052712 (2025).

\bibitem{CDWBe25}
M. W. C. Dharma-wardana and D. D Klug, Phys. Rev. E {\bf 111}, 065208 (2025).

\bibitem{cdwSi20}
M.W.C. Dharma-wardana, Dennis D. Klug, and Richard C. Remsing
Phys. Rev. Lett. {\bf 125}, 075702 (2020). 
doi: 10.1103/PhysRevLett.125.075702



\bibitem{CDW-Pool25}
M. W. C. Dharma-wardana, D. D. Klug, Hannah. Poole, and G. Gregori.
Phys. Rev. E {\bf 111}(1) 015205 (2025).


\bibitem{Hull20}
Hull, C. J.,  Raj, S. L., \&. Saykally, R. J.
 Chemical Physics Letters {\bf 749}, 137341 (2020).




\bibitem{cdw-carb22}
M. W. C. Dharma-wardana and Dennis D. Klug,
 Phys. Plasmas {\bf 29}, 022108  (2022); doi: 10.1063/5.0077343

\bibitem{VASP}
G. Kresse and J. Furthm\"{u}ller, Phys. Rev. B {\bf 54}, 11169 (1996).
\url{https://www.vasp.at}

\bibitem{ABINIT}
X. Gonze, B. Amadon, G. Antonius, F. Arnardi, et. al.
Computer Physics Communications 2 {\bf 248}, 107042. (2020).
\url{https://doi.org/10.1016/j.cpc.2019.107042}.

\bibitem{KarasievHu19}
 V. V. Karasiev, S. X. Hu, M. Zaghoo, and T. R. Boehly, ?Exchange-correlation
thermal effects in shocked deuterium: Softening the principal hugoniot and
thermophysical properties. Phys. Rev. B {\bf 99}, 214110 (2019).

 

\bibitem{Ramakrishna20}
 K. Ramakrishna, T.Dornheim, and Jan Vorberger.
Phys. Rev. B {\bf 101}, 195129 (2020);
Erratum, {\bf 112}, 119901 (2025).

\bibitem{KarasievQE14}
V. V. Karasiev, T. Sjostrom, S. B. Trickey, Computer Physics Communications,
{\bf 185} 2340-3249 (2014).


\bibitem{Recou05}
 V. Recoules and J. P. Crocombette, Phys. Rev. B {\bf 72},
104202 (2005).

\bibitem{MoldHyd25}
Zhandos A. Moldabekov, Xuecheng Shao, Hannah M. Bellenbaum, 
Cheng Ma, Wenhui Mi, Sebastian Schwalbe, Jan Vorberger, and Tobias Dornheim.
arXiv:2507.00688v1 [physics.plasma-ph] (2025).



\bibitem{Plage-XRTS15}  %
K-U Plageman, H. R. R\"{u}ter, T. Bornath, Mohammed Shihab,
Michael P. Desjarlais, C. Fortmann, S. Glenzer, R. Redmer.
 Phys. Rev. E {\bf 92}, 013103 (2015).


\bibitem{xrt-Harb16}
L. Harbour, M. W. C. Dharma-wardana, D. Klug and L. Lewis. 
Physical Review E {\bf 94}, 053211, (2016).



\bibitem{SternZbar07} 
P.A. Sterne S.B. Hansen, B.G. Wilson, W.A. Isaacs, 
High Energy Density Phys. {\bf 3}, 278 (2007).

\bibitem{BethkenZbar20}
Mandy Bethkenhagen, Bastian B. L. Witte, Maximilian Sch\"{o}rner,
Gerd R\"{o}pke, Tilo D\"{o}ppner,
 Dominik Kraus, Siegfried H. Glenzer, Philip A. Sterne, and Ronald Redmer.
Phys. Rev. Research {\bf 2}, 023260  (2020).



\bibitem{PerdZung81}
J. P. Perdew and Alex Zunger
Phys. Rev. B {\bf 23}, 5048 (1981).

\bibitem{cdw-Carbon10E6-21}
M.W.C. Dharma-wardana, 
Ionization of carbon at 10-100 times the diamond density and in
 the 10$^6$ K temperature range.
Phys. Rev. E {\bf 104}, 015201 (2021).


\bibitem{Kohn86}
W. Kohn Phys. Rev. B {\bf 33}, 4331(R)  (1986).


\bibitem{Fauss21}
Faussurier, G., Blancard, C. \&  Bethkenhagen, M.
Phys. Rev. E {\bf 104}, 025209 (2021).

\bibitem{Sham85}
L. J. Sham and Schl\"{u}ter, Phys. Rev. B {\bf 32}, 3883 (1985).
Density functional theory of the bandgap.


\bibitem{HybLouie85}
M. A. Hybertson and S. G. Louie, Phys. Rev. Lett., {\bf 55}, 1418 (1985).

\bibitem{PDW-Dyson84}
F. Perrot and M. W. C. Dharma-wardana, Phys. Rev. A {\bf 29}, 1378 (1984).

\bibitem{Trickey23}
W. Mi, K. Luo, S. B. Trickey, M. Pavonello, 
Chemical Reviews, {\bf 123}, 12039 (2023).




\bibitem{Loh90}
E. Y. Loh, J. E. Gubernatis, R. T. Scalettar, S. R. White, D. J. Scalapino, and R. L. Sugar.
Phys. Rev. B {\bf 41}, 9301?9307 (1990).


\bibitem{Troyer05}
M. Troyer and U. J. Wiese, 
 Phys. Rev. Lett. {\bf 94},170201 (2005).

\bibitem{Dornheim2019}
T. Dornheim, 
 Phys. Rev. E {\bf 100}, 023307 (2019).


\bibitem{DornheimPIMC24}   %
T. Dornheim, T. D\"{o}ppner, P. Tolias,
M. P. B\"{o}hme, L.B. Fletcher, Th. Gawne, F. R. Graziani,
D. Kraus, M. J. MacDonald, Zh. A. Moldabekov,
S. Schwalbe, D.O. Gericke, and J. Vorberger.
arXive:2402.19113v1 [physics.plasmas-ph]  (2024).



\bibitem{driver12}
K. P. Driver, 
B. Militzer,
 Phys. Rev.  Lett. \textbf{108}, 115502 (2012).

\bibitem{ZhangMilitzeCH-18}
Shuai Zhang,
Burkhard Militzer,
Lorin X. Benedict,
Fran\'{c}ois Soubiran,
Philip A. Sterne,
and Kevin P. Driver
J. Chem. Phys.,  {\bf 148} (10), 102318  (2018).



\bibitem{Brown2013}
Ethan W. BrownJ. L. DuboisJ. L. Dubois, Markus Holzmann, David Ceperley
Phys. Rev B {\bf 88}, 081102(R) (2013).


\bibitem{GDS17}
S. Groth, T. Dornheim, T. Sjostrom, F.D. Malone, W. Foulkes, M. Bonitz,
Phys. Rev. Lett. {\bf 119} (13)  135001 (2017).
http://dx.doi.org/10.1103/PhysRevLett.119.135001.




\bibitem{DWP82}
M. W. C. Dharma-wardana and F. Perrot. 
Phys. Rev. A {\bf 26}, 2096  (1982).

\bibitem{Furutani90}
 F. Perrot, Y. Furutani and M.W.C. Dharma-wardana,
Phys. Rev. A {\bf 41}, 1096-1104 (1990).


\bibitem{ChiharaNPA91}
J. Chihara, Phys. Rev. A {\bf 41}, 1247 (1991).


\bibitem{Pe-Be} 
 F. Perrot,  Phys. Rev. E {\bf 47}, 570 (1993).

\bibitem{Hungary16}
M. W. C. Dharma-wardana,
Proceedings of the Conference on Density Functional Theory, Debrecen, 2015.
 Edited by K. Schwarz and A. Nagy. 
Computation  {\bf 4} (2), 16; (2016). 

\bibitem{ELR98}
M. W. C. Dharma-wardana, and Fran\c{c}ois Perrot, 
Phys. Rev. E {\bf 58}, 3705 (1998).

\bibitem{DW-yuk22}
M. W. C. Dharma-wardana, Lucas J. Stanek, and Michael S. Murillo
Phys. Rev. E {\bf 106}, 065208 (2022).



\bibitem{HamelCH12}
S. Hamel, 
Lorin X. Benedict, Peter M. Celliers, M. A. Barrios,
 T. R. Boehly, G. W. Collins, Tilo D\"{o}ppner,  J. H. Eggert, 
D. R. Farley, D. G. Hicks, J. L. Kline, A. Lazicki, S. LePape, A. J. Mackinnon,
 J. D. Moody, H. F. Robey, Eric Schwegler, and Philip A. Sterne,
Phys.  Rev. B {\bf 86}, 094113 (2012).

\bibitem{whitley15}
H. D. Whitley, 
D. M. Sanchez , S. Hamel , A. A. Correa,
and L. X. Benedict, Contrib. Plasma Phys. {\bf 55}, 390 (2015).


\bibitem{Stanek21}
 Lucas J. Stanek, Raymond C. Clay III, M. W. C. Dharma-wardana, 
Mitchell A. Wood, Kristian R. C. Beckwith, and Michael S. Murillo,
Phys. Plasmas {\bf 28}, 032706 (2021).


\bibitem{cdw-pop21}
M. W. C. Dharma-wardana, 
Physics of Plasmas {\bf 28}, 052108 (2021); https://doi.org/10.1063/5.0047642
Simple pair-potentials and pseudo-potentials for warm-dense matter and general applications. 

\bibitem{GaneshSiLPPT-09}
 P. Ganesh, and M. Widom, Phys. Rev. Lett. {\bf 102}, 075701
(2009).

\bibitem{Remsing20}
Richard C. Remsing and Michael L. Klein,
J. Phys. Chem 2020
 https://dx.doi.org/10.1021/acs.jpcb.0c01798




\bibitem{PDWXC}
F. Perrot and M. W. C. Dharma-wardana, Phys. Rev. B {\bf 62}, 16536 (2000);
{\it Erratum: } {\bf 67}, 79901 (2003); arXive-1602.04734.



\bibitem{Filinov2004}
A. V. Filinov, V. O. Golubnychiy, M. Bonitz, W. Ebeling, 
and J. W. Dufty. Phys. Rev. E {\bf 70}, 046411 (2004).





\bibitem{prl1}
M. W. C. Dharma-wardana and F. Perrot, Phys. Rev. Lett. {\bf 84}, 959 (2000).


\bibitem{prl2}
Fran\c{c}ois Perrot and M. W. C. Dharma-wardana,  Phys. Rev. Lett. {\bf 87},
 206404 (2001).

\bibitem{SandipDufty13}
Dufty, J.; and Dutta, Sandipan;  Phys. Rev. E. {\bf 87}, 032101 (2013).


\bibitem{lfc1-dw19}
M. W. C. Dharma-wardana,
D. Neilson and F. M. Peeters
Phys. Rev. B {\bf 99}, 035303 (2019).


\bibitem{Bredow15}
R. Bredow, Th. Bornath, W.-D. Kraeft, M.W.C. Dharma-wardana and R. Redmer
Contributions to Plasma Physics, 
 {\bf 55}, 222-229 (2015).
 DOI 10.1002/ctpp.201400080


\bibitem{cdw-N-rep19}
M. W. C. Dharma-wardana, Phys. Rev. B {\bf 100}, 155143 (2019).
DOI: 10.1103/PhysRevB.100.155143


\bibitem{LiuWuCHNC14}
Yu Liu and Jianzhong Wu,
J. Chem. Phys {\bf 141} 064115 (2014).



\bibitem{hug02}
Dharma-wardana, M. W. C.;  and  Perrot, F.;
  Phys. Rev. B  {\bf  66}, 014110 (2002).

\bibitem{Norman68}
G. E. Norman, A. N. Starostin, High Temp.  {\bf 6}, 394 (1968).


\bibitem{Ashcroft68}
N. W. Ashcroft, Metallic hydrogen: A high-temperature superconductor?
Phys. Rev. Lett. {\bf 21}, 1748-1749 (1968).

\bibitem{Murillo13}
M. S. Murillo, 
 J. Weisheit, S. B. Hansen, and M. W. C. Dharma-wardana,
Phys. Rev. E {\bf 87}, 063113 (2013).

\bibitem{Gawne24}
Thomas Gawne, Sam M. Vinko,
 and Justin S. Wark. Phys. Rev. E {\bf 103}, L023201 (2024).

\bibitem{Sharma25}   
V. Sharma and A. J. White, Phys. Rev. Lett. {\bf 134}, 095102 (2025).

\bibitem{PironBlenski11}
R. Piron and T. Blenski, Phys. Rev. E {\bf 83}, 026403 (2011).

\bibitem{SXHu16}
S. X. Hu, L. A. Collins, V. N. Goncharov, J. D. Kress, R. L. McCrory, S. Skupsky.
Physics of Plasmas, {\bf 23}, 042704 (2016).

\bibitem{Pain2023}
Nadine Wette and J-C Pain, Phys. Rev. E {\bf 108}, 015205 (2023).

\bibitem{LFA83}
F. Lado, S. M. Foiles and N. W. Ashcroft,  Phys. Rev. {\bf A 26}, 2374 (1983).


\bibitem{Ebeling85}
W. Ebeling, and W. Richert. Physics Letters A, {\bf 108} 80 (1985).

\bibitem{SaumonChab89}
D. Saumon, and G. Chabrier.  Phys. Rev. Lett. {\bf 62}, 2397 (1989).

\bibitem{Weir96}
(1992).
S. T. Weir, A. C. Mitchell, and W. J. Nellis, Metallization of fluid molecular
hydrogen at 140 GPa (1.4 Mbar), Phys. Rev. Lett. {\bf 76}, 1860-1863 (1996).

\bibitem{Margo96}
W. R. Magro, D. M. Ceperley, C. Pierleoni, and B. Bernu, Molecular dissocia-
tion in hot, dense hydrogen, Phys. Rev. Lett. {\bf 76}, 1240-1243 (1996).


\bibitem{Filinov2001}
V. Filinov, V. Fortov, M. Bonitz, and P. Levashov, Phase transition in strongly
degenerate hydrogen plasma, Jetp. Lett. {\bf 74}, 384 (2001).

\bibitem{Filinov2003}
V. Filinov, M. Bonitz, P. Levashov, V. Fortov, W. Ebeling, and M. Schlanges,
Plasma phase transition in hydrogen and electron-hole plasmas, Contrib.
Plasma Phys. {\bf 43}, 290-294 (2003).



\bibitem{NormanSait17}
G. E. Norman and I. M. Saitov.
 Doklady Physics,  {\bf 62}, No. 6, pp. 284-288. (2017).
Original Russian, G.E. Norman, I.M. Saitov,Doklady Akademii Nauk, {\bf 474} (5) 553?557 (2017).
 
\bibitem{MilitzerCataldo21}
B. Militzer, F. Gonzalez-Cataldo, S. Zhang, K. P. Driver, and F. Soubiran,
Phys. Rev. E 103, 013203 (2021).

\bibitem{Holst11}
B. Holst, M. French, and R. Redmer, Electronic transport coefficients from ab
initio simulations and application to dense liquid hydrogen, Phys. Rev. B {\bf 83},
235120 (2011).


\bibitem{prl3}
M. W. C. Dharma-wardana  and F. Perrot,
 Phys. Rev. Lett.  {\bf 90}, 136601 (2003).

\bibitem{Jost05}
D. Jost and M. W. C. Dharma-wardana
Phys. Rev. B {\bf 72}, 195315 (2005).

\bibitem{Totsuji09}
Chieko Totsuji, Takashi Miyake, Kenta Nakanishi,
Kenji Tsuruta and Hiroo Totsuji
J. Phys.: Condens. Matter {\bf 21}  045502 (2009).


\bibitem{Chihara2000}
 J. Chihara, J. Phys.: Condens. Matter {\bf12}, 231 (2000).
\bibitem{FanoRau86}
U. Fano and A. R. P. Rau, Atomic collisions and spectra. New York, Academic. (1986).

\bibitem{Friedel52}
J. Friedel, Philosophical Magazine, {\bf 43}, 153 (1952).

\end{thebibliography}
\end{document}